\documentstyle[sprocl1]{article}
\input{psfig}
\def\Journal#1#2#3#4{{#1} {\bf #2}, #3 (#4)}

\def\NPB{{\em Nucl.~Phys.}~B}

\def\PRD{{\em Phys.~Rev.}~D}
\def\be{\begin{equation}}
\def\ee{\end{equation}}
\def\bea{\begin{eqnarray}}
\def\eea{\end{eqnarray}}
\def\VEV#1{\langle #1\rangle}
\def\gsim{\mbox{\raisebox{-.6ex}{~$\stackrel{>}{\sim}$~}}}

\begin{document}

\title{ELECTROWEAK BARYOGENESIS IN\\
 THE MINIMAL SUPERSYMMETRIC STANDARD MODEL}

\author{ J.M.~CLINE }

\address{McGill University, Department of Physics,
Montr\'eal, Qu\'ebec H3A 2T8, Canada}

\maketitle\abstracts{I describe work done in collaboration with
M.~Joyce and K.~Kainulainen on (1) the strength of the electroweak
phase transition in the MSSM and (2) the mechanism for producing the
baryon asymmetry during the phase transition.  In the former we compare
the effective potential and dimensional reduction methods for
describing the phase transition and search the parameter space of the
MSSM for those values where it is strong enough.  In the latter we give
a systematic computation of the baryon asymmetry due the CP-violating
force acting on charginos in the vicinity of the bubble wall.  We find
that a light right-handed stop, a light Higgs boson, and a large phase
in the $\mu$ parameter, are the main necessary ingredients for
producing the baryon asymmetry.}

\section{Experimentally testable baryogenesis?} A major triumph of
astroparticle physics is our understanding of the light element
abundances through big bang nucleosynthesis.  Equally exciting is the
prospect of putting the measured baryon asymmetry of the universe,
$\eta = n_B/n_\gamma\cong 3\times 10^{-10}$, on a similar footing.  To
do so we must have a theory that will soon be testable in the
laboratory.  The minimal supersymmetric standard model (MSSM) provides
such a framework.

Let us first review Sakharov's three necessary ingredients for
baryogenesis. (1) Baryon number violation: this is already present in
the standard model (SM) in the form of $\Delta B=3$ sphaleron
interactions, involving 9 left-handed quarks and 3 leptons.  (2) CP
violation: this is too weak for baryogenesis in the SM, but potentially
strong enough in the MSSM via two independent phases, which can be
taken to be those of the $\mu$ term and the soft-breaking $A$ terms:
$|\mu|e^{i\delta_\mu} H_1 H_2$ in the superpotential, and $y_t |A_t|
e^{i\delta_A}\tilde q_L h_2\tilde t_R^c$ in the soft-breaking part of
the potential.  (3) Loss of thermal equilibrium: the electroweak phase
transition can be first order, going by nucleation of bubbles of the
new vacuum with nonvanishing Higgs field VEV's, $\VEV{H}= v_c $, in the
symmetric vacuum where $\VEV{H}= 0$.  If the transition is strong
enough, the sphaleron interactions will be slower than the expansion
rate of the universe inside the bubbles, which is necessary for
preventing the baryon asymmetry from relaxing to zero.  Outside the
bubbles, the sphaleron rate is fast, which is necessary to create the
baryon asymmetry in the first place.  At the critical temperature
$T_c$, the sphaleron rates go like $\Gamma_{\rm sph} \sim
\alpha_W^{4-5?} T_c$ outside the bubbles and $e^{-c v_c/T_c}$ inside,
where $c$ is a constant, and the exponent of $\alpha_W$ has been
recently called into question.\cite{pa} To insure that $\Gamma_s$ is
small enough inside the bubbles, one finds the necessary condition
$v_c/T_c > 1$.

The two crucial questions for successful electroweak baryogenesis are
thus (1) under what conditions is $v_c/T_c >1$, so that baryon washout
is avoided?  And (2) do parameters exist such that a large enough
baryon asymmetry can be created in the first place?

\section{Strength of the phase transition: when is ${\bf v_c/T_c >1}$?}
I will review two methods that have been used to compute the critical
ratio $v_c/T_c$.  Traditionally one used the finite-temperature
effective potential (EP), which in the $\overline{MS}$ scheme is the
familiar expression\vskip-0.2in
\bea
  V_{\rm eff} &=&  V_{\rm tree}(H_i)
  + {\textstyle \frac{1}{64\pi^2} 
  \sum_i \pm m_i^4(H) \left(\ln {m^2_i(H)\over Q^2}-\frac32
  \right)}\nonumber\\
	&+&
{\textstyle \sum_i \pm T \int{ d^3p\over(2\pi)^3}\ln\left(1\pm e^{-\sqrt{
	p^2 + m_i^2(H)}/T}\right)}\nonumber,
\eea \vskip-0.1in\noindent
plus higher order corrections, like the important resummation of
ring-diagrams.  Here the sum is over all particles, with their masses
evaluated at arbitrary backgound Higgs VEV's $H_i$.  A problem with the
EP is that some of these masses (the transverse gauge bosons) are very
small when $H_i\sim 0$, giving rise to infrared divergences in the
momentum integral.  The cost of adding an extra loop of transverse $W$'s
to an arbitrary loop diagram in the EP is of order $g^2 T/m_W(H)$.
%\be
%	g^2 T \int{ d^3p \over (2\pi)^3}{p^2\over(p^2+m_W^2(H))^3} 
%	\sim {g^2 T\over m_W(H)}
%\ee
These same contributions are responsible for giving the hump in the
Higgs potential that makes the transition first order.  Thus the
potential suffers from potentially big uncertainties in just the region
where one would like to know it best to accurately gauge the strength
of the transition.

A way around this problem is dimensional reduction (DR),\cite{laine}
where one computes an effective theory by integrating out all {\it
but} these most dangerous infrared contributions, leaving the lattice
to take care of them.   For example, a typical bosonic loop
contribution of the form $\int {d^4p\over 2\pi^4} (p^2+m^2)^{-2}$ at
$T=0$ would become a sum over Matsubara frequencies,
$T\sum_n\int{d^3p\over 2\pi^3}(\vec p^{\,2} + (2n\pi T)^2 +m^2)^{-2}$ at
finite temperature.  Only the terms with $n=0$ can diverge as $m\to 0$
because the others have a large effective mass, $2n\pi T$.  In DR, an
effective three dimensional Lagrangian of the $n=0$ modes is obtained
which has the form ${\cal L}_{\rm eff} = |\vec D H|^2 +
(1/4g_3^2)F_{ij}F_{ij} -\mu^2_3|H|^2 + \lambda_3|H|^4$, where $H$ is
the light linear combination of the two Higgs doublets of the MSSM.  It
has been found that the limit $v_c/T_c>1$ corresponds to
$x_c\equiv\lambda_3/g_3^2< 0.044$.  Thus one need only compute $x_c$ in
terms of the parameters of the MSSM to find which ones are consistent
with baryogenesis.

We have searched the MSSM parameter space using both the EP and DR to
compare the two approaches,\cite{ck} and found that they are roughly
compatible, though DR gives a somewhat bigger range of allowed
parameters, as shown in figure 1.  The most sensitive parameters along
with their predicted values are $\tan\beta\equiv \VEV{H_2}/{H_1} =
2^{+1.5}_{-0.5}$, $m_{\rm higgs} < 85$ GeV, a light stop with $\tilde
m_- = 150-200$ GeV, and a correspondingly small right-handed soft
breaking stop mass parameter of $m_U< 100$ GeV.  Negative values of
$m_U^2$ have been advocated for making the transition strong enough,
but our results show that they are not necessary.  Furthermore we find
that arbitrarily large values of $\mu$ are allowed, which is helpful
because $\mu$ is the major source of CP violation in the model.  This
is in contrast to previous studies that have emphasized the weakening
effect of $\mu$ on the transition.

\begin{figure}
\centerline{\psfig{figure=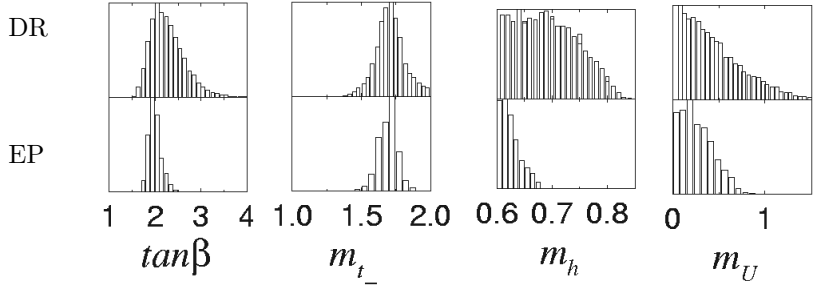,height=1.5in}}
\caption{Distributions of parameters satisfying the baryogenesis phase 
transition
constraint in the MSSM, using dimensional reduction (top line) and the
effective potential (bottom line).  Masses are in units of 100 GeV.}
\vskip-2in
\leftline{DR}
\vskip0.5in
\leftline{EP}
\vskip 1in
\end{figure}

\section{Origin of the baryon asymmetry---classical force mechanism}
Knowing the preferred parameters for safeguarding the baryon asymmetry,
we now ask whether it can be generated in sufficient quantity in the
first place.  Several studies of electroweak baryogenesis in the MSSM
have been carried out,\cite{ewb} some invoking simplifying assumptions
we find questionable. Our derivation\,\cite{cjk} starts from first
principles, showing clearly how CP violation in the wall becomes a
source for the Boltzmann equations for the transport of chirality in
front of the wall.

The physical picture is quite intuitive.  Squarks, quarks, and
charginos (the charginos turn out to dominate) undergo interactions as
they pass through the expanding wall, where they experience
CP-violating forces and scatterings, and this leads to a
buildup of chirality in front of the wall, for example, an excess of
$\tilde h_{2,L}$'s over their antiparticles.  The excess of chirality in a
given species is transmitted to the left-handed quarks through top
Yukawa interactions, and this biases sphalerons to produce baryon
number that eventually falls inside the bubbles.  The origin of the
CP-violating effects on charginos is the $\mu$ parameter which appears
in the chargino mass term,\newline
\centerline{
$\left(\begin{array}{cc} \overline{\tilde W}_R^+ & 
	\overline{\tilde h}^+_{1,R} \end{array}\right)
\left(\begin{array}{cc} \tilde m_2 & g v_2(z)/\sqrt{2}\\
	g v_1(z)/\sqrt{2} & |\mu|e^{i\delta} \end{array} \right)
	\left(\begin{array}{cc} \tilde W_L^+ & \tilde h^+_{2,L}
	\end{array}\right)$.}
%\bea
%	&\left(\begin{array}{cc} \overline{\tilde W}_R^+ & 
%	\overline{\tilde h}^+_{1,R} \end{array}\right)
%	\left(\begin{array}{cc} \tilde m_2 & g v_2(z)/\sqrt{2}\\
%	g v_1(z)/\sqrt{2} & |\mu|e^{i\delta} \end{array} \right)
%	\left(\begin{array}{cc} \tilde W_L^+ & \tilde h^+_{2,L}
%	\end{array}\right)&\nonumber
%\eea

Since the wall is much thicker ($w\sim 20/T$) than the average thermal
wavelength of the particles $(1/T)$, it makes sense to treat Higgs
VEV's as slowly varying.  Thus we diagonalize the mass matrix locally
at each point in the wall, and solve the Dirac equation using the WKB
approximation.  One finds in this way that the mass eigenvalues have a
spatially varying complex phase, $\pm\theta(z)$, which gives rise to a
difference between their dispersion relations, as well as those for
particles and antiparticles, in the form $p(z) = \sqrt{E^2-m^2(z)}
\pm{\rm sign}(p_z)(m/2E)^2\partial_z\theta$.  This is the origin of
the CP violating force.  In addition, when the interactions are
rewritten in terms of the local mass eigenstates, they acquire
dependence on $\theta(z)$, revealing CP violation in the scatterings.
Not only these effects but also the scatterings that change charginos
into quarks and squarks are incorporated into the Boltzmann equations
in a well-defined manner, leading to diffusion equations of the form
$-D_i n''_i -v_w n'_i +\sum_j\Gamma_{ij}n_j = S_i\sim v_w m^2\theta'$
for the $i$th species, where $D_i$ is the diffusion coefficient, $v_w$
the wall velocity, and $S_i$ the source term which can be
computed unambiguously.  One can solve these equations for the
left-handed quark density $n_L$, from which the baryon asymmetry
follows using $n_B = \int \dot n_B dt\sim (\Gamma_{\rm sph}/v_w)
\int_0^\infty n_L dz$.  The quantity to be compared to determinations
from big bang nucleosynthesis is the baryon-to-photon ratio, $\eta =
n_B/n_\gamma\cong 3\times 10^{-10}$.

The final answer for the baryon asymmetry depends on $\tilde m_2$ (the
Wino mass parameter), $\mu$, $\tan\beta$, the Higgs and quark diffusion
constants, the wall velocity $v_w$, the strong and weak sphaleron
rates, the rate of top Yukawa interactions, the wall width, and the
critical VEV and temperature. It is also proportional to the underlying
CP violation in the $\mu$ parameter, $\sin\delta$.  Putting in our best
estimates for these quantities (or the ranges allowed by the constraint
$v_c/T_c>1$) we find that typically $\sin\delta$ must be unity to get a
large enough asymmetry, although if $\tilde m_2\cong\mu$ then it is
possible to have $\sin\delta \sim 0.1$.  The dependence of
$\eta_{10}=10^{10}\eta$ on $\mu$ and $\tilde m_2$ is shown for several
choices of other parameters in figure 2.

\begin{figure}
\centerline{\psfig{figure=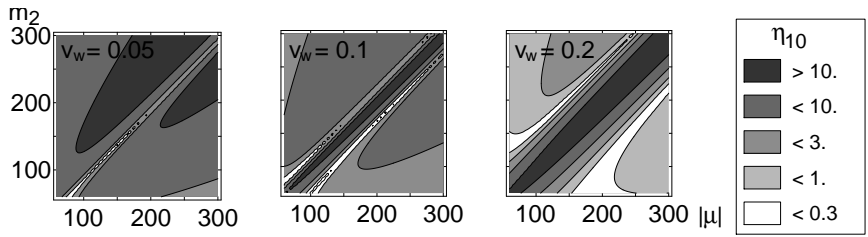,height=1.21in}}
\caption{Contours of constant $\eta_{10}=10^{10}\eta$ in the plane of 
$\tilde m_2$ (the Wino mass) and $\mu$, in GeV, $v_{\rm wall}= 0.05$, $0.1$ 
and $0.2$, $\tan\beta=2$, $\sin\delta=1$ and wall width $= 14/T$.}
\end{figure}

Such large values of the phase $\delta$ as we require are often
considered to be incompatible with bounds on the neutron electric
dipole moment, but this is only true if both the squarks and the
charginos are relatively light.  If the up and down squarks are much
heavier than the stop and sbottom, the EDM bound on $\delta$ is
relaxed.  In fact there exist scenarios where such a hierarchy between
third and lower generation squarks arises naturally.  

\section{Summary} It appears that producing the baryon asymmetry in the
MSSM is not easy; the parameter space is highly constrained.  From the
perspective seeking an experimentally falsifiable theory of
baryogenesis, this is a encouraging.  The phase transition is strong
enough only with a relatively light stop and higgs, and $\tan\beta\sim
2$.  Baryon production is maximized when $\mu\sim\tilde m_2$, hence the
Wino and Higgsino are roughly degenerate.  Also the CP violation in the
$\mu$ parameter must satisfy $\sin\delta\gsim 0.1$, which indicates a
large neutron EDM, in fact too large unless the lower generation
squarks are much heavier than the light stop.  If supersymmetry should
be confirmed in the next few years and these predictions verified, it
will be very tempting to believe we finally have evidence for the true
origin of the baryon asymmetry of the universe.


\section*{References}
\begin{thebibliography}{99}
\bibitem{pa} P.~Arnold, in these proceedings, hep-ph/9706305 (1997).
\bibitem{laine} M.~Laine, in these proceedings, hep-ph/9707415 (1997).
\bibitem{ck}J.M.~Cline and K.~Kainulainen, \Journal{\NPB}{482}{73}{1996},\\
hep-ph/9605235; preprint McGILL-97-7, hep-ph/9705201 (1997).
\bibitem{ewb} P.~Huet and A.E.~Nelson, \Journal{\PRD}{53}{4578}{1996},
hep-ph/9506477;\\
M.~Aoki, N.~Oshimo and A.~Sugamoto, hep-ph/9612225 (1996);\\
M.~Carena, M.~Quiros, A.~Riotto, I.~Vilja and C.E.M.~Wagner,
preprint CERN-TH-96-242, hep-ph/9702409 (1997).
\bibitem{cjk}J.M.~Cline, M.~Joyce and K.~Kainulainen, ``Supersymmetric
Electroweak Baryogenesis in the WKB Approximation," preprint
McGill-97-26, hep-ph/9708393 (1997).

\end{thebibliography}
\end{document}